\documentclass[allclo]{FBSart}
\usepackage{amsfonts}
\usepackage{amssymb}
\usepackage{epsfig}
\usepackage{bm}
\def\Vec#1{\mbox{\boldmath $#1$}}
\def\eqne{\end{equation}}
 
\def\eqnb{\begin{equation}}

\def\NPB{{Nucl. Phys.} {\bf B}}
\def\PLB{{Phys. Lett.} B}

\def\PRD{{Phys. Rev.} D}
\def\PRC{{Phys. Rev.} C}

\newcommand{\Slash}[1]{\ooalign{\hfil/\hfil\crcr$#1$}}

\title{Propagator of the lattice domain wall fermion and the staggered fermion}
\author{Sadataka Furui}
\institute{School of Science and Engineering, Teikyo University.\\
1-1 Toyosatodai, Utsunomiya, 320-8551 Japan\thanks{\textit{E-mail address:} furui@umb.teikyo-u.ac.jp }}

\runningauthor{Sadataka Furui}
\runningtitle{Propagator of the lattice domain wall fermion and the staggered fermion}
\begin{document}

\maketitle
\begin{abstract}
We calculate the propagator of the domain wall fermion (DWF) of the RBC/UKQCD collaboration with 2+1 dynamical flavors of $16^3\times 32\times 16$ lattice in Coulomb gauge, by applying the conjugate gradient method. We find that the fluctuation of the propagator is small when the momenta are taken along the diagonal of the 4-dimensional lattice. 
Restricting momenta in this momentum region, which is called the cylinder cut, we compare the mass function and the running coupling of the quark-gluon coupling $\alpha_{s,g_1}(q)$ with those of the staggerd fermion of the MILC collaboration in Landau gauge. 

In the case of DWF, the ambiguity of the phase of the wave function is adjusted such that the overlap of the solution of the conjugate gradient method and the plane wave at the source becomes real. The quark-gluon coupling $\alpha_{s,g_1}(q)$ of the DWF in the region $q>1.3$GeV agrees with ghost-gluon coupling $\alpha_s(q)$ that we measured by using the configuration of the MILC collaboration, i.e. enhancement by a factor $(1+c/q^2)$ with $c\simeq 2.8$GeV$^2$ on the pQCD result.

In the case of staggered fermion, in contrast to the ghost-gluon coupling $\alpha_s(q)$ in Landau gauge which showed infrared suppression, the quark-gluon coupling $\alpha_{s,g_1}(q)$ in the infrared region increases monotonically as $q\to 0$.  Above 2GeV, the quark-gluon coupling $\alpha_{s,g_1}(q)$ of staggered fermion calculated by naive crossing becomes smaller than that of DWF, probably due to the complex phase of the propagator which is not connected with the low energy physics of the fermion taste. 

\end{abstract}


\maketitle

\section{Introduction}
In the calculation of quark-gluon vertices in the infrared region, non-perturbative renormalization is possible by calculating the quark propagator in a fixed gauge.  The calculation of the quark propagator on lattice is reviewed in \cite{BHLWZ05}.  In our previous paper \cite{FuNa05}, we studied the quark propagator of staggered fermion in Landau gauge using the full QCD configurations of relatively large lattice ($24^3\times 64$) of MILC collaboration \cite{MILC} available from the ILDG data base \cite{ILDG2}. In the last year, full QCD configurations of the domain wall fermion (DWF) of medium size ($16^3\times 32\times 16$)  were released in the ILDG and in this year large size ($24^3\times 64\times 16$) were released \cite{ILDG1} from the RBC/UKQCD collaboration \cite{AABB07}.

In these configurations the length of the 5th dimension was fixed to be 16.  In this paper we show the results of the medium size DWF configurations and compare with the results of the large size staggered fermion.

Charcteristic features of infrared QCD are confinement and chiral symmetry breaking. The confinement is related to the Gribov copy i.e. non gauge uniqueness, which makes the sharp evaluation of physical quantities difficult, and we try to fix the gauge in the fundamental modular region \cite{Zw94}.  Chiral symmetry breaking is speculated to be related to instantons \cite{tH86}. The Orsay group discussed that the infrared suppression of the triple gluon coupling is due to instantons\cite{Orsay01, Orsay02}.
 The running coupling from the quark-gluon coupling in quenched approximation also showed similar infrared behavior \cite{Sk97}.  Our simulation of the ghost-gluon coupling in Landau gauge obtained by configurations of the MILC collaboration showed infrared suppression, but in Coulomb gauge the running coupling $\alpha_I(q)$ of MILC and of RBC/UKQCD did not show suppression \cite{FuNa07}. Thus, it is interesting to check the difference of the Coulomb gauge and the Landau gauge, staggered fermion and DWF, and the ghost-gluon coupling and the quark-gluon coupling.

The domain wall fermion (DWF) was first formulated by Kaplan in 1992 \cite{Ka92,Ka93}
by assuming that the chiral fermion couples with the gauge field in the fifth dimension.
The model was improved by Narayanan and Neuberger \cite{NaNe93} and Shamir \cite{Sha93a,Sha93b}, such that the gauge field are strictly four dimensional and are copied to all slices in the fifth dimension. The model was applied in the finite temperature simulation of $8^3\times 4$ lattice with $L_s$ from 8 to 32 lattices \cite{CCFKM01} and to quenched simulation of $8^3\times 32, 12^3\times 32$, and $16^3\times 32$ lattices with $L_s$ from 16 to 64 \cite{BCCCD04}. 

 The fermionic part of the Lagrangian formulated for the lattice simulation is \cite{Sha93a, Sha93b, FurSha94}, 
\begin{equation}
S_F(\bar\psi, \psi,U)=-\sum_{x,s;y,s'}\bar\psi_{x,s}(D_F)_{x,s;y,s'}\psi_{y,s'},
\end{equation}
where
\begin{equation}
(D_F)_{x,s;y,s'}=\delta_{s,s'}D^\parallel_{x,y}+\delta_{x,y}D^\perp_{s,s'}.\label{D_F}
\end{equation}
The interaction $D^\parallel$ contains the gauge field and the interaction in the fifth dimension defined by $D^\perp$ does not contain the gauge field\cite{BCCCD04}.

The bare quark operators are defined on the wall at $s=0$ and $s=L_s-1$ as
\begin{equation}
q_x=P_L \psi_{x,0}+P_R \psi_{x,L_s},
\end{equation}
where $\displaystyle P_R=\frac{1+\gamma_5}{2}$ and $\displaystyle P_L=\frac{1-\gamma_5}{2}$ are the projection operator.
For the Dirac's $\gamma$ matrices, we adopt the convention of ref. \cite{CCFKM01}, in which $\gamma_5$ is diagonal.

In the DWF theory, a Lagrangian density in the fermion sector 
\begin{equation}
\mathcal{L}_1=i\bar\psi(\Slash\partial -i\Slash A)\psi+\bar\psi(MP_R+M^\dagger P_L)\psi
\end{equation}
with an operator $M$ acting on the left-handed and right-handed field was proposed \cite{NaNe93}.  
In this method, the free fermion propagator becomes
\begin{eqnarray}
&&[\Slash{p}-M^\dagger P_L-M P_R]^{-1}=(\Slash{p}+M)P_L\frac{1}{p^2-M^\dagger M}\nonumber\\
&&+(\Slash{p}+M^\dagger)P_R\frac{1}{p^2-MM^\dagger}.
\end{eqnarray}

On the lattice, one introduces the bare quark mass $m_f$ that mixes the two chiralities, and 5 dimensional mass $M_5$.
The lattice simulation of DWF propagator is performed by introducing a Hamiltonian, whose essential idea is given in Appendix.
This formalism was adopted in the Schwinger model \cite{Vra98} and in the 4-dimensional lattice simulation \cite{CCFKM01, ABBB06, AABB07}.

Instead of using the transfer matrix method, we calculate the quark propagator by using the conjugate gradient method in five dimensional spaces. We interpret the configuration at the middle of the two domain walls in the fifth dimension as the physical quark wave function.
 We measure the propagator of the domain wall fermion using the configurations of the RBC/UKQCD, and compare the propagator with that of the configurations of MILC \cite{FuNa05}. We measure also the quark-gluon coupling from the quark propagator, by applying the Ward identity.

The organization of this paper is as follows. In sect. 2, we present a formulation of the lattice DWF and its numerical results are shown in sect.3.
In sect.4 the lattice calculation of the staggered fermion propagator which we adopted in \cite{FuNa05} is summarized and in sect. 5, a comparison of the DWF fermion-gluon and the staggered fermion-gluon is given. 
Conclusion and discussion are given in the sect.6.  Some comments on the Hamiltonian is given in the Appendix.

\section{The lattice calculation of the DWF propagator}

In this section we present the method of calculating the DWF propagator.

Using the $D_F$ defined in eq.\ref{D_F}
we make a hermitian operator $D_H=\gamma_5 R_5 D_F$, where $(R_5)_{ss'}=\delta_{s,L_s-1-s'}$ is a reflection operator as
\begin{eqnarray*}D_H=&&\left(\begin{array}{cccccc} -m_f\gamma_5 P_L &  & & & \gamma_5 P_R & \gamma_5(D^\parallel -1)\\& & & \gamma_5P_R & \gamma_5({D^\parallel}-1)& \gamma_5 P_L \\   & & \cdots&\cdots & \cdots&   \\ & & \cdots &\cdots &\cdots &  \\ \gamma_5P_R & \gamma_5({D^\parallel}-1)& \gamma_5 P_L & & &\\ \gamma_5({D^\parallel}-1)& \gamma_5 P_L & & & & -m_f\gamma_5 P_R \end{array}\right)\\
\end{eqnarray*}
where $P_{R/L}=(1\pm\gamma_5)/2$, and -1 in $(D^\parallel -1)$ originates from $D^\perp_{s,s'}$.

The quark sources are sitting on the domain walls as
\begin{equation}
q(x)=P_L\Psi(x,0)+P_R\Psi(x,L_s-1).
\end{equation}
We take $P_L\Psi(x,s)\propto e^{-(\frac{s}{r})^2}$, $P_R\Psi(x,s)\propto e^{-(\frac{L_s-s-1}{r})^2}$, $(s=0,1, \cdots, L_s-1)$ with $r=0.842105$, which corresponds to $(r/L_s)/(1+r/L_s)=0.05$ \cite{deGLo91}.

  In the case of quenched approximation, a condition on the Pauli-Villars regularization mass  $M$ for producing a single fermion with the left hand chirality bound to $s=0$ and the right bound to $s=L_s-1$ is $0<M<2$. In the free theory, the condition that the transfer matrix along the 5th dimension be positive yields  a restriction $0<M<1$ \cite{NaNe93}. However, in a quenched interacting system,  $M=1.8$ was adopted \cite{BCCDF02} and since in the conjugate gradient method  there is no $M<1$ constraint, we adopt the same value.

We define the base of the fermion as
\begin{equation}
\Psi(x)={^t(}\phi_L(x,0),\phi_R(x,0),\cdots,\phi_L(x,L_s-1),\phi_R(x,L_s-1))\\
\end{equation}
where $^t$ means the transpose of the vector, and the
$\phi_{L/R}(x,l_s)$ contains color $3\times 3$ matrix, spin $2\times 2$ matrix
and $n_x\times n_y\times n_z\times n_t$ site coordinates.  We measure $9\cdot 4$ matrix elements on each site at once.

The wave functions $\phi_{L/R}(x,s)$ are solutions of the equation 
\begin{eqnarray}
&&\gamma_5({D^\parallel}-1)\left(\begin{array}{c}\phi_L(x,s)\\
                                         \phi_R(x,s)\end{array}\right)\nonumber\\
&&=\left(\begin{array}{cc} 1&0\\
                        0&-1\end{array}\right)
\left(\begin{array}{cc}-B & C\\
                       -C^\dagger& -B\end{array}\right)\left(\begin{array}{c}\phi_L(x,s)\\
                                         \phi_R(x,s)\end{array}\right)\nonumber\\
&&=\left(\begin{array}{cc} -B& C\\
                           C^\dagger & B\end{array}\right)\left(\begin{array}{c}\phi_L(x,s)\\
                                         \phi_R(x,s)\end{array}\right)
\end{eqnarray}
where 
\begin{equation}
B=(5-M_5)\delta_{xy}-\frac{1}{2}\sum_{\mu=1}^4(U_\mu(x)\delta_{x+\hat\mu,y}+{U^\dagger}_\mu(y)\delta_{x-\hat\mu,y}),
\end{equation}
and 
\begin{equation}
C=\frac{1}{2}\sum_{\mu=1}^4(U_\mu(x)\delta_{x+\hat\mu,y}-{U^\dagger}_\mu(y)\delta_{x-\hat\mu,y})\sigma_\mu.
\end{equation}
Here
\begin{equation}
M_5=M \theta(s-L_s/2)=\left\{ \begin{array}{cc} -M & s<\frac{L_s-1}{2}\\
                                    M & s\geq \frac{L_s-1}{2}\end{array}\right. \end{equation}
is the mass introduced for the regularization.
Using the $\gamma$ matrices as defined in \cite{CCFKM01}, we obtain
\begin{eqnarray}
C(x,y)P_R&=&\frac{1}{2}\sum_{\mu=1}^4 (U_\mu(x)\delta_{x+\hat\mu,y}-{U^\dagger}_\mu(y)\delta_{x-\hat\mu,y})\Sigma P_R,\nonumber\\
C^\dagger(x,y)P_L&=&\frac{1}{2}\sum_{\mu=1}^4 ({U^\dagger}_\mu(x)\delta_{x+\hat\mu,y}-{U}_\mu(y)\delta_{x-\hat\mu,y})\Sigma^\dagger P_L\nonumber\\
\end{eqnarray}
where
\[
\Sigma=\left(\begin{array}{c} i\sigma_1\\
                                                -i\sigma_2\\
                                                i\sigma_3\\
                                                -I\end{array}\right)\quad
{\rm and}\quad
\Sigma^\dagger=\left(\begin{array}{c} -i\sigma_1\\
                                                i\sigma_2\\
                                                -i\sigma_3\\
                                                -I\end{array}\right).
\]

The conjugate gradient method for solving the 5 dimensional DWF propagator is a  simple extension of the method we used in the staggered fermion \cite{FuNa05}, since the degrees of freedom in the 5th dimension can be treated as if they are internal degrees of freedom on each 4 dimensional sites.

As in the transfer matrix method, we define
\begin{equation}
\bar M=\left(I+\frac{1}{5-M_5}D_H\right),
\end{equation}

\begin{equation}
L=\left( \begin{array}{cc}0 & 0 \\
                         -\frac{1}{5-M_5}D_{H\, eo}& 0\end{array}\right)
\end{equation}
and
\begin{equation}
U=\left(\begin{array}{cc}0 &  -\frac{1}{5-M_5}D_{H\, oe}\\0& 0\end{array}\right)
\end{equation}
such that
\begin{equation}
(1-L)^{-1}\bar M(1-U)^{-1} 
=\left(\begin{array}{cc}I & 0\\       
                0 & I-\frac{1}{5-M_5}D_{H\, eo}D_{H\, oe}\end{array}\right),
\\
\end{equation}
where even-odd decomposition is done in the 5 dimensional space. We solve the equation for
\[
\phi=\left(\begin{array}{c}\phi'_o\\ \phi'_e\end{array}\right)
\]
using the source
\[
\frac{1}{5-M_5}\rho=\rho'=\left(\begin{array}{c} \rho'_o\\    \rho'_e\end{array}\right),
\]
\begin{equation}
\left(I-\frac{1}{(5-M_5)^2}D_{H\, eo}D_{H\, oe}\right)\phi_e= \rho'_e-\frac{1}{5-M_5}D_{H\, eo}\rho'_o.
\end{equation}

The solution on the odd sites is calculated from that of even sites as
\begin{equation}
\phi_o=\rho'_o-\frac{1}{5-M_5}D_{H oe}\phi'_e.
\end{equation}

In the process of conjugate gradient iteration, we search shift parameters for $\alpha_k^L$ \cite{FuNa04} for $\phi_L$ and $\alpha_k^R$ for $\phi_R$ and in the first 50 steps we choose $\alpha_k=Min(\alpha_k^L,\alpha_k^R)$ and shift $\phi_{k+1}^L=\phi_k^L-\alpha_k\phi_k^L$ and $\phi_{k+1}^R=\phi_k^R-\alpha_k\phi_k^R$ and in the last 25 steps we choose $\alpha_k=Max(\alpha_k^L,\alpha_k^R)$, so that the stable solution is selected for both $\phi_L$ and $\phi_R$.

The convergence condition attained in this method is about $0.5\times 10^{-4}$. One can improve the condition by increasing the number of iteration, but the overlap of the solution and the plane wave do not change significantly.

To evaluate the propagator, we measure the trace in color and spin space of the inner product in the momentum space between the plane waves
\[
\chi(p)={^t(}\chi_L(p,0),\chi_R(p,0),\cdots,\chi_L(p,L_s-1),\chi_R(p,L_s-1))
\]
and the solution of the conjugate gradient method
\[
\Psi(p)={^t(}\phi_L(p,0),\phi_R(p,0),\cdots,\phi_L(p,L_s-1),\phi_R(p,L_s-1))
\]
as 
\[
{\rm Tr} \langle \bar\chi(p,s) P_L \Psi(p,s)\rangle=Z_B(p)(2N_c) {\mathcal B}_L(p,s),
\]
\begin{equation}
{\rm Tr} \langle \bar\chi(p,s) P_R \Psi(p,s)\rangle=Z_B(p)(2N_c) {\mathcal B}_R(p,s)
\end{equation}
and
\[
{\rm Tr} \langle \bar\chi(p,s) i\Slash{p}P_L\Psi(p,s)\rangle=Z_A(p)/(2N_c) i{\bf p}{\mathcal A}_L(p,s),
\]
\begin{equation}
{\rm Tr} \langle \bar\chi(p,s) i\Slash{p}P_R\Psi(p,s)\rangle=Z_A(p)/(2N_c) i{\bf p}{\mathcal A}_R(p,s)
\end{equation}
where $\displaystyle p_i=\frac{1}{a}\sin \frac{2\pi \bar p_i}{N_i}$ ($\bar p_i=0,1,2,\cdots,N_i/2$).

On the lattice at each $s$ the 4-dimensional torus is residing. We perform the Fourier transform in the 4-dimensional space, but take the momentum in the 5th direction to be zero since it corresponds to the lowest energy state. $Z_A(p)$ and $Z_B(p)$ are the wave function renormalization factor.

When $p_4=0$, the term ${\mathcal B}(p,s)$ are given by the matrix elements of $\langle \chi_R,\Psi_L\rangle$ and $\langle \chi_L, \Psi_R\rangle$.
The operator $\Slash{p}$ yields matrix elements of $\langle\chi,\Sigma \Psi_L\rangle$ and $\langle\chi,\Sigma \Psi_R \rangle$. The propagator is parametrized as
\begin{equation}
S(p)=[\frac{-i\Slash{p}+{\mathcal M}^\dagger(\hat p)}{p^2+{\mathcal M}(\hat p){\mathcal M}^\dagger(\hat p)}P_L]+[\frac{-i\Slash{p}+{\mathcal M}(\hat p)}{p^2+{\mathcal M}^\dagger(\hat p){\mathcal M}(\hat p)}P_R]
\end{equation}
where
\[
{\mathcal M}(\hat p)=\frac{Re[{\mathcal B}_R(p,L_s/2)]}{Re[{\mathcal A}_R(p,L_s/2)]}
\]
and \[
{\mathcal M}^\dagger(\hat p)=\frac{Re[{\mathcal B}_L(p,L_s/2)]}{Re[{\mathcal A}_L(p,L_s/2)]}.
\]
The momentum assignment $\displaystyle \hat p_i=\frac{2}{a}\sin\frac{\pi \bar p_i}{N_i}$ is 
introduced for removing doublers using the Wilson prescription.

The ${\mathcal M}(\hat p)$ has zero eigenfunction and $dim(Ker{\mathcal M})=n_R=1$
and the ${\mathcal M^\dagger}(\hat p)$ does not have zero eigenfunction and $dim(Ker{\mathcal M^\dagger})=n_L=0$.

\section{Numerical results of the DWF propagator}
The configurations of the RBC/UKQCD collaboration are first Landau gauge fixed and then Coulomb gauge fixed ($\partial_i A_i=0$) as follows\cite{FuNa07}.
We adopt the minimizing function 
$F_U[g]=||{\Vec A}^g||^2=\sum_{x,i}{\rm tr}
 \left({{A^g}_{x,i}}^{\dag}A^g_{x,i}\right)$,
and solve $\partial_i ^gA_i({\Vec x},t)=0$ using the Newton method. We obtain $\displaystyle \epsilon=\frac{1}{-\partial D}\partial_i {A_i}$ from the eq. $\partial_i A_i+\partial _i D_i(A)\epsilon=0$.  Putting $g({\Vec x},t)=e^\epsilon$ in ${U^g}_i({\Vec x},t)=g({\Vec x},t)U_i({\Vec x},t)g^\dagger({\Vec x+i},t)$ we set the ending condition of the gauge fixing as the maximum of the divergence of the gauge field over $N_c^2-1$ color and the volume is less than $10^{-4}$,
$
Max_{x,a}(\partial_i A_{x,i})^a <10^{-4}
$.
This condition yields in most samples
\[
\frac{1}{8V}\sum_{a,x}(\partial_i {A_{x,i}^a})^2\sim 10^{-13}.
\]
We leave the remnant gauge on $A_0(x)$ unfixed, but since the Landau gauge preconditioning is done, it is not completely random.  We leave the problem of whether a random gauge transformation, or the remnant gauge fixing on $A_0(x)$ modify the propagators, but we do not expect drastic corrections will happen.

Using the gauge configurations of RBC/UKQCD collaboration after Coulomb gauge fixing, we calculate ${\rm Tr}\langle \chi(p,s) \phi_L(p,s)\rangle$ and ${\rm Tr}\langle \chi(p,s)i\Slash{p} \phi_L(p,s)\rangle$  and  ${\rm Tr}\langle \chi(p,s) \phi_R(p,s)\rangle$ and ${\rm Tr}\langle \chi(p,s)i\Slash{p} \phi_R(p,s)\rangle$ at each 5-dimensional slice $s$.  
Number of samples is 49 for each mass $m_f=0.01$, 0.02 and 0.03. We measured
in certain momentum directions of $m_f=0.01$, 149 samples.

In our Lagrangian there is a freedom of choosing global chiral angle in the 5th direction,
\begin{equation}
\psi \to e^{i\eta \gamma_5}\psi, \qquad \bar\psi \to \bar \psi e^{-i\eta\gamma_5}\psi.
\end{equation}

We adjust this phase of the matrix element such that both ${\rm Tr}\langle \chi(p,0) \phi_L(p,0)\rangle$ and ${\rm Tr}\langle \chi(p,L_s-1) \phi_R(p,L_s-1)\rangle$ are close to a real number. Namely, we define 
\[
e^{i\theta_L}=\frac{{\rm Tr}\langle \chi(p,0) \phi_L(p,0)\rangle}{|{\rm Tr}\langle \chi(p,0) \phi_L(p,0)\rangle|},
\]
\[
e^{-i\theta_R}=\frac{{\rm Tr}\langle \chi(p,L_s-1) \phi_R(p,L_s-1)\rangle}{|{\rm Tr}\langle \chi(p,L_s-1) \phi_R(p,L_s-1)\rangle|}
\]
and sample-wise calculate $e^{i\eta}$ such that
\begin{equation}
|e^{i\theta_L}e^{i\eta}+1|^2+|e^{i\theta_R}e^{-i\eta}-1|^2
\end{equation}
is minimum.   When $\bar p$ is even and the momentum is not along
the diagonal of the four dimensional system, we also calculate $e^{i\eta'}$ such that
\begin{equation}
|e^{i\theta_L}e^{i\eta'}-1|^2+|e^{i\theta_R}e^{-i\eta'}-1|^2
\end{equation}
is minimum, but the final results by multiplying $e^{i\eta}$ and $e^{i\eta'}$ are similar.

In the calculation of $\mathcal B_{L/R}$, we define matrix elements multiplied by the phase as
\[
\widetilde{\langle \chi(p,s)\phi_L(p,s)\rangle}=\langle \chi(p,s)\phi_L(p,s)\rangle e^{-i \eta},
\]
\[
\widetilde{\langle \chi(p,s)\phi_R(p,s)\rangle}=\langle \chi(p,s)\phi_R(p,s)\rangle e^{i \eta}.
\]
and correspondingly denote ${\mathcal B}_{L/R}(p,s)$ multiplied by the phase $e^{-i\eta}$ and $e^{i\eta}$ as $\tilde{\mathcal B}_{L/R}(p,s)$, respectively.

 In the calculation of ${\mathcal A}_{L/R}(p,s)$, we diagonalize
\begin{eqnarray}
&&[\widetilde{\langle \chi(p_x,s)\phi_L(p_x,s)\rangle} \sigma_1+
\widetilde{\langle \chi(p_y,s)\phi_L(p_y,s)\rangle} \sigma_2\nonumber\\
&&+\widetilde{\langle \chi(p_z,s)\phi_L(p_z,s)\rangle} \sigma_3+
\widetilde{\langle \chi(p_z,s)\phi_L(p_z,s)\rangle}i I]
\end{eqnarray}
and
\begin{eqnarray}
&&[\widetilde{\langle \chi(p_x,s)\phi_R(p_x,s)\rangle }\sigma_1+
\widetilde{\langle \chi(p_y,s)\phi_R(p_y,s)\rangle} \sigma_2\nonumber\\
&&+\widetilde{\langle \chi(p_z,s)\phi_R(p_z,s)\rangle} \sigma_3
+\widetilde{\langle \chi(p_z,s)\phi_R(p_z,s)\rangle} i I],
\end{eqnarray}
 where $I$ is the $2\times 2$ diagonal matrix, and define the 
  ${\mathcal A}_{L/R}(p,s)$ multiplied by the phase 
 as $\tilde{\mathcal A}_{L/R}(p,s)$.

The term ${\mathcal B}_{L/R}$ is a sum of color-spin diagonal scalar, while the term ${\mathcal A}_{L/R}$ is a color-diagonal but momentum dependent spinor and  we take the positive eigenvalue.  

In order to minimize the artefact due to violation of rotational symmetry of the lattice we restrict the momentum configuration to be diagonal in the 4-d lattice. This prescription which is called cylinder cut \cite{BBLWZ01} is already adopted in ghost propagator \cite{FuNa06a} and in quark propagator \cite{FuNa05, BHLWZ05} calculations.

In general, there is a mixing between $\phi_L$ and $\phi_R$ and there
is a sign problem i.e. the sign of $Re[\tilde{\mathcal B}_{L/R}(p,L_s/2)]$ and $Re[\tilde{\mathcal A}_{L/R}(p,L_s/2)]$ becomes random. The sign is related to the sign of the source at $s=0$ and $s=L_s-1$. But, when the cylinder cut is chosen, the sign problem does not seem to occur.

In the calculation of the propagator of DWF, the mass originates not only from the mid-point matrix $Q^{(mp)}$ defined as
\begin{equation}
{Q^{(mp)}}_{s,s'}=P_L\delta_{s,L_s/2}\delta_{s',L_s/2}+P_R\delta_{s,L_s/2-1}\delta_{s',L_s/2-1} \label{Qmp}
\end{equation}
but also from $Q^{(w)}$ defined as
\begin{equation}
{Q^{(w)}}_{s,s'}=P_L\delta_{s,0}\delta_{s',0}+P_R\delta_{s,L_s-1}\delta_{s'L_s-1}
\end{equation}

At zero momentum the numerator ${\mathcal B}_L(p=0,s=0)$ becomes 1 and it gives a contribution of $m_fQ^{(w)}=m_f$. Since there is no pole mass in $\phi_R(s,l_s)$, the value of ${\mathcal B}_R(p=0,s=L_s-1)$ is not physical.
In the midpoint contribution $\displaystyle \frac{Re[\tilde{\mathcal B}_{L/R}(p,L_s/2)]}{Re[\tilde {\mathcal A}_{L/R}(p,L_s/2)]}$, we take into account that the numerator of the mass function contains $(2N_c)\times (2N_c)$ coherent contributions and divide by the multiplicity.

\begin{figure}[htb]
\begin{center}
\epsfig{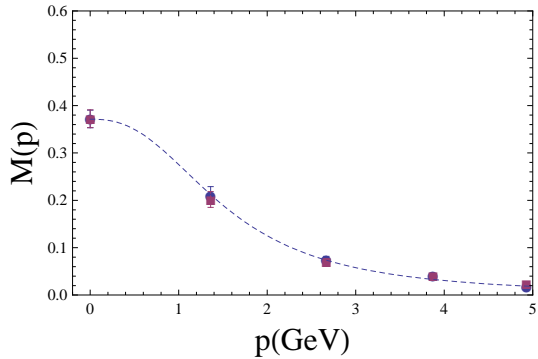}
\caption{The mass function in GeV of the domain wall fermion as a function of the modulus of Euclidean four momentum $p$(GeV). $m_f=0.01$. (149 samples). Blue disks are $m_L$ (left handed quark) and red boxes are $m_R$ (right handed quark).}
\label{dwf_mass1}
\end{center}
\end{figure}

The Fig.\ref{dwf_mass1} is the mass function of $m_f=0.01/a=0.017$GeV. 
The momenta correspond to $\bar p=(0,0,0,0),(1,1,1,2),(2,2,2,4),(3,3,3,6)$ and $(4,4,4,8)$. The dotted lines are the
phenomenological fit 
\begin{equation}
{\mathcal M}(\hat p)=\frac{c\Lambda^{2\alpha+1}}{p^{2\alpha} + \Lambda^{2\alpha}} + \frac{m_f}{a}
\end{equation}
Since the pole mass $Q^{(w)}$ is not
included in the plots, $m_f$ is set to be 0 here. The corresponding values of $0.02/a=0.034$GeV and $0.03/a=0.050$GeV are similar. 

In the $\chi^2$ fit, we choose $\alpha$ equals 1,1.25 and 1.5 and searched best values for $c$ and $\Lambda$. We found the global fit is best for $\alpha=1.25$.  
The fitted parameters are given in Table \ref{massparameter}. 
\begin{table}
\begin{center}
\begin{tabular}{cccccc}
 &$m_{ud}/a$ & $m_s/a$ & $c$ &  $\Lambda$(GeV) & $\alpha$ \\
\hline
DWF$_{01}$ &0.01   &0.04 &  0.24  &  1.53(3)  & 1.25 \\
DWF$_{02}$ &0.02  &0.04  &  0.24  &  1.61(5)  & 1.25 \\
DWF$_{03}$ &0.03  &0.04  &  0.30  &  1.32(4)  & 1.25 \\
\hline
MILC$_{f1}$ &0.006 &0.031  &  0.45 & 0.82(2) & 1.00 \\
MILC$_{f2}$ &0.012 &0.031  &  0.43 & 0.89(2) & 1.00 \\
\hline
\end{tabular}
\end{center}
\caption{The fitted parameters of mass function of DWF(RBC/UKQCD) and staggered fermion (MILC) with the staple plus Naik action.}\label{massparameter}
\end{table}

In the case of staggered fermion in Landau gauge, we adopt in this work the staple plus Naik action on $m_f=0.0136$GeV and 0.027GeV configurations\cite{FuNa05,FuNa06}. We fixed the parameter $\alpha=1$ and obtained $\Lambda=0.82$GeV and 0.89GeV, respectively. In general $\Lambda$ becomes larger for larger $\alpha$, but $\Lambda$ of RBC/UKQCD seems larger than that of MILC, which is also observed in the quenched overlap fermion propagator \cite{Bon02}. In the case of MILC, $\Lambda$ becomes smaller for smaller mass $m_f$, but in the case of RBC/UKQCD, it is opposite.
Analytical expression of the quark propagator in Dyson-Schwinger equation is formulated in \cite{RW94, AvS01} and a comparison with these lattice data are given in \cite{ADFM04, BPRT03}.
The mass function of the staggered fermion is close to that of the Dyson-Schwinger equation of $N_f=3$, but larger than that of the $N_f=0$.
\begin{figure}[htb]
\begin{center}
\epsfig{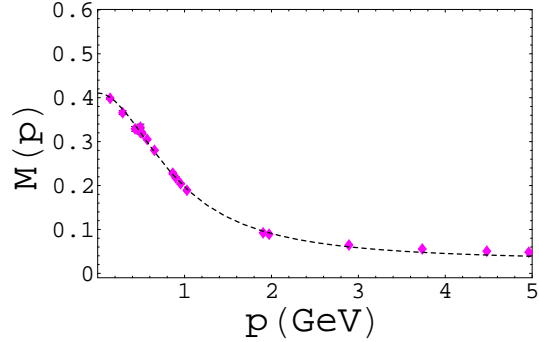}
\caption{The mass function in GeV of the staggered fermion MILC$_{f2}$ as a function of the modulus of Euclidean four momentum $p$(GeV). The staple plus Naik action is adopted }\label{milc_mass2}
\end{center}
\end{figure}

The error bars are taken from the Bootstrap method after 5000 re-samplings \cite{whit90,var}.
The re-sampling method reduces the error bar by about a factor of 10 as compared to the standard deviation of the bare samples.

We measured also momentum points $(\bar p,0,0,0)$, $(0,\bar p,0,0)$, $(0,0,\bar p,0)$ and $(\bar p,\bar p,\bar p,0)$ with
$\bar p=1,2,3$ and 4, but the error bars of the mass function are found to be large especially at (2,2,2,0).
 We observed systematic difference of the magnitude of mass functions of $(\bar p,0,0,0)$ with even $\bar p$ and odd $\bar p$. Such problems would be solved by improving statistics, and systematically correcting deviation from the spherical symmetry, but at the moment these problems can be evaded by adopting the cylinder cut.

\section{The lattice calculation of the staggered fermion propagator} 
The lattice results of fermion propagator using staggered (Asqtad) action and overlap action are reviewed in \cite{BHLWZ05}.
We calculated the propagator of staggered fermion of the MILC collaboration using the conjugate gradient method \cite{FuNa05}.
The inverse quark propagator is expressed as
\begin{equation}
S_{\alpha\beta}^{-1}(p,m)=i\sum_\mu(\bar\gamma_\mu)_{\alpha\beta}\left[\frac{9}{8}\sin(p_\mu)-\frac{1}{24}\sin(3p_\mu)\right]+m\bar\delta_{\alpha\beta}
\end{equation}
where $\alpha_\mu=0,1$, $\beta_\mu=0,1$ and $\bar\delta_{\alpha\beta}=\prod_\mu\delta_{\alpha_\mu\beta_\mu|mod 2}$. The momentum of the staggered fermion
$k_\mu$ takes values $k_\mu=p_\mu+\pi\alpha_\mu$ where
\begin{equation}
p_\mu=\frac{2\pi m_\mu}{L_\mu}, \quad m_\mu=0,\cdots,\frac{L_\mu}{2}-1.
\end{equation}

The $\gamma$ matrix of staggered fermions is 
\begin{equation}
(\bar\gamma_\mu)_{\alpha\beta}=(-1)^{\alpha_\mu}\bar\delta_{\alpha+\zeta^{(\mu)},\beta}
\end{equation}
where 
\begin{equation}
\zeta_\nu^{(\mu)}=\left\{\begin{array}{ll}1& {\rm if } \quad\nu<\mu\\
                                          0& {\rm otherwise}\end{array}\right.
\end{equation}
The ${\mathcal A}(p)$ is defined as
\begin{equation}
i\sum_\alpha\sum_\mu(-1)^{\alpha_\mu} p_\mu Tr[\sum_\beta S_{\alpha\beta}(p)]=16N_c p^2{\mathcal A}(p)
\end{equation}
where 16 is the number of taste.

The staggered fermion incorporating the lattice symmetries including parity and charge conjugation by introducing a general mass matrix is formulated in \cite{GS84}. In our model, we do not incorporate the general mass matrix, but take the
same mass as MILC collaboration.

  The chiral symmetry of the staggered fermion of the MILC collaboration is currently under discussion \cite{Cz07,BGSS07,Ad08}. In our simple model, the charge conjugation operator can be taken as $C=-i\gamma_4$ and $C\gamma_\mu^TC^\dagger=-\gamma_\mu$.  We interpret $p$ in the original as $q$ in crossed channel and since staggered actions are invariant under translation of 2$a$, we modify the scale by a factor of 1/2:
\begin{equation}
p=\frac{1}{2a}\left[\frac{9}{8}\sin(p_\mu)-\frac{1}{24}\sin(3p_\mu)\right]
\end{equation}
where $\displaystyle \frac{1}{a}=2.19$GeV/c and 2.82 GeV/c in the MILC$_{f1}$ and MILC$_{f2}$ respectively.

\section{A comparison of the DWF-gluon and the staggered fermion-gluon coupling}
In this section we calculate the running coupling in the crossed channel $q\bar q\to$ gluon of the DWF and compare with that of the staggered fermion.
In Coulomb gauge, the quark gluon vertex from the three point Green function becomes \cite{SK02,Orsay03}
\begin{equation}
G_\mu(p,q)=\int d^4x\int d^4 y e^{ipy-i(p+q)x}\langle \psi(y)\bar\psi(0)\gamma_\mu\psi(0)\bar\psi(x)\rangle.
\end{equation} 
When the momentum transfer $\Vec q$ is small, the vertex function satisfying the
Ward identity $\displaystyle Z_V\Gamma_\mu(p)=-i\frac{\partial}{\partial p_\mu}S^{-1}(p)$ becomes
\begin{eqnarray}
\Gamma_\mu(p,q)&=&S^{-1}(p)G_\mu(p,q)S^{-1}(p)\nonumber\\
&=&\delta^{ab}[g_1(p^2)\gamma_\mu+ig_2(p^2)q_\mu+g_3(p^2)p_\mu \Slash{q}]
\nonumber\\
\end{eqnarray}
where $\delta^{ab}$ is the delta function in color space and $Z_V$ is the vertex renormalization factor. 

When the contribution of the ghost-quark coupling \cite{EF74} is ignored, the vector current Ward identity allows us to extract the running coupling $\alpha_{s,g_1}(q)$ from the difference of $S^{-1}({\Vec p}+\frac{\Vec q}{2})$ and $S^{-1}({\Vec p}-\frac{\Vec q}{2})$ \cite{MPSTV94}
\begin{equation}
-i[S^{-1}(({\Vec p}+{\frac{\Vec q}{2}})_j|0)-S^{-1}(({\Vec p}-{\frac{\Vec q}{2}})_j|0)]=Z^V\Lambda_{0}(p) \frac{{\Vec q}_j}{4\pi}.\\
\end{equation}

When the crossing is performed, the momentum transfer becomes
$\displaystyle \Vec p+\frac{\Vec q}{2}-(-\Vec p+\frac{\Vec q}{2})=2\Vec p$. 
For massless fermion with cylinder cut, $4{\Vec p}^2=p^2$ and thus $p$ can be
interpreted as $q$. 

In the case of DWF, we diagonalize
\begin{equation}
\sum_{j=1}^3[\langle{{\mathcal A}_L}^{\alpha\beta}({\Vec p}+\frac{\Vec q}{2})_j-{{\mathcal A}_L}^{\alpha\beta}({\Vec p}-\frac{\Vec q}{2})_j\rangle\sigma_j]
\end{equation}
and
\begin{equation}
\sum_{j=1}^3[ \langle{{\mathcal A}_R}^{\alpha\beta}({\Vec p}+\frac{\Vec q}{2})_j-{{\mathcal A}_R}^{\alpha\beta}({\Vec p}-\frac{\Vec q}{2})_j\rangle\sigma_j], 
\end{equation}
and to get the running coupling, we evaluate the average and multiply the normalization $\displaystyle Z^V(p)\propto \frac{Z^2}{E(p)}=\frac{(2N_c)^4}{2E(p)}\times \frac{1}{2}$ where  $\displaystyle\frac{1}{2E(p)}$ is from
normalization of the $\mathcal B={\mathcal M}{\mathcal A}$ in the original
direct channel, which should be proportional to $\displaystyle\frac{\mathcal M}{2E(p)}$ 
and $\displaystyle\frac{1}{2}$ comes from fixing the incoming wave as $q$ or $\bar q$ and
$Z=(2N_c)^2$ comes from the relative normalization of ${\mathcal A}$ and
$\mathcal B$.

\begin{figure}[htb]
\begin{center}
\epsfig{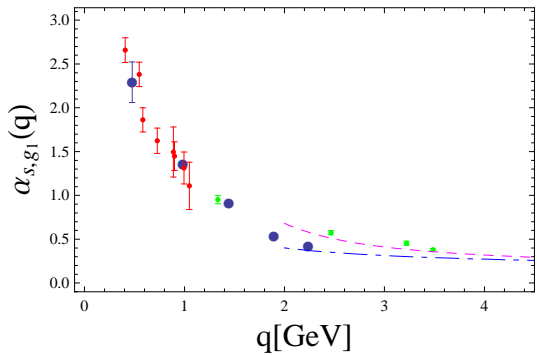}
\caption{The running coupling $\alpha_{s,g_1}(q)$ of MILC$_{f1}$ (blue disks) and DWF$_{01}$ (green points). The dash-dotted line is the pQCD result and the dashed line is the pQCD with the $A^2$ condensate contribution. The red points are data extracted from the experiment by the JLab group.}\label{alpha}
\end{center}
\end{figure}

In Fig.\ref{alpha}, we show the running coupling of DWF$_{01}$ and MILC$_{f1}$.  An enhancement of $\alpha_{s,g_1}(q)$ of RBC/UKQCD above 2GeV region could be the effect of the $A^2$ condensate due to instantons \cite{Orsay02}.  
Using operator product expansion, the Orsay group fitted the lattice data above 2.6Gev as
\begin{equation}
\alpha_s^{Latt}(q^2)=\alpha_{s,pert}(q^2)(1+\frac{c}{q^2})
\end{equation}
where the parameter $c$ is proportional to the $A^2$ condensate. They
obtained $\displaystyle c=2.7(1.2)\left[\frac{a^{-1}(\beta=5.6,\kappa_{sea}=0.1560)}{2.19{\rm GeV}} {\rm GeV}\right]^2$.

 We show pQCD result with $N_f=3$ without (dot-dashed line) and with the $A^2$ condensate effect (dashed line) where $c=2.8$GeV$^2$ is used \cite{FuNa06b} as in the analysis of the ghost-gluon coupling. Consequence of the condensates on the mass gap and quark confinement is discussed in \cite{KMSI02}.

In $q>2$GeV, the running coupling $\alpha_{s,g_1}(q)$ of MILC$_{f1}$ is smaller than that of RBC/UKQCD. We think it is due to the complex phase of the staggered fermion in the ultraviolet region which is not related to the low energy physics of the fermion taste \cite{Ad08}. In the previous analysis of propagator of the MILC$_{f2}$ in which the bare $s$-quark mass is close to the bare $u/d$ quark mass,  we observed an anomalous behavior when the Asqtad action is adopted \cite{FuNa06}. Since the sample size was not large, we cannot exclude the possibility that the anomalous behavior of the staggered fermion disappear in the simulation of larger number of samples. We leave these problems in the future.

The running coupling in the infrared region is consistent with the experimental data extracted by the JLab group \cite{DBCK06}. They compared the proton form factor and the neutron form factor and by adopting the Drell-Hearn-Gerasimov sum rule in infrared and the Bjorken sum rule in ultraviolet, extracted the running coupling.

\section{Conclusion and discussion}
The mass function and the running coupling in Coulomb gauge of the gauge configuration of RBC/UKQCD collaboration were calculated and compared with those of the staggered fermion. 
We adopted the conjugate gradient method and imposed a reality condition on the overlap of the distorted wave and the plane wave at the position of the fermion sources. 

We observed that the quark-gluon coupling $\alpha_{s,g_1}(q)$ of the DWF is consistent with the ghost-gluon coupling $\alpha_s(q)$ of MILC$_{f1}$ in $q > 1.3$GeV region. 
The running coupling $\alpha_{s,g_1}(q)$ of the staggered fermion
in Landau gauge (MILC$_{f1}$) does not show infrared suppression, in contrast to the ghost-gluon coupling \cite{FuNa06b}.
The discrepancy of the ghost-gluon coupling $\alpha_s(q)$ and the quark-gluon coupling $\alpha_{s,g_1}(q)$ in Landau gauge suggests that there is a problem in the ghost-gluon coupling, and/or the color structure of the loop given by a product of ghosts \cite{F08, F08a}.  The difference of Landau gauge quark-gluon vertex in quenched configuration \cite{Sk97} and in unquenched configuration that we measured, suggests important contribution of fermions in the dynamics of gluons. Orsay group interpreted the infrared suppression of triple gluon vertex of quenched configuration as the instanton effect \cite{Orsay03}. In the infrared, however, the vacuum amplitude with presence of instantons contain  zero-mode divergence \cite{tH86}. In the expression of the triple gluon vertex
\begin{equation}
\alpha_s(q)=\frac{1}{4\pi}\left[\frac{G^{(3)}(q^2,q^2,q^2)}{(G^{(2)}(q^2))^3}(q^2G^{(2)}(q^2))^{3/2}\right]
\end{equation}
it was assumed that $\displaystyle G^{(3)}(q^2,q^2,q^2)\propto \frac{n}{48p}\langle \rho^9I(q\rho)^3\rangle$ and $\displaystyle G^{(2)}(q^2)=\frac{n}{8}\langle \rho^6I(q\rho)^2\rangle$, $n$ being the instanton density and $\rho$ being the instanton radius. When $\rho$ is large, the zero-mode divergence could overwhelm $q^4$ dependence and yields a constant $\alpha_s(q)$. 

In a supersymmetric theory, the zero-mode divergence from fermion and from  boson are shown to cancel out \cite{AdVe77}.  
In quenched simulation of gluonic systems, the fermionic zero mode divergence is absent, and consequently incorrect large $\rho$ dependence of instantons could have introduced the infrared suppression of the triple gluon running coupling.

Although we do not consider the supersymmetric Yang-Mills theory, we show in Appendix that $M$ and ${M}^\dagger$ could be regarded as supersymmetric interactions.
 Phenomenologically, the cancellation of zero-mode divergence from quark field and that from gluon field in the conjugate gradient calculation of the quark propagator seem to have introduced the correct infrared behavior.

The new method of deriving the quark propagator is encouraging, but it is necessary to extend the calculation to larger lattice for getting the continuum limit, and to extend the simulation for other momenta that are far from the 4-dimentional diagonal axis. The origin of the fluctuation of the data outside the cylinder cut region is under investigation.

\vskip 0.2 true cm
\begin{acknowledge}
The author thanks Reinhard Alkofer for a discussion on the quark propagator in
Coulomb gauge and the support of author's stay in Graz in March 2008, and
Hideo Nakajima for the collaboration in the early stage of this project and producing the gauge fixed configurations.

The numerical simulation was performed on Hitachi-SR11000 at High Energy Accelerator Research Organization(KEK) under a support of its Large Scale Simulation Program (No.07-04 and No.08-01), and on NEC-SX8 at Yukawa institute of theoretical physics of Kyoto University.
\end{acknowledge}
\appendix
\section{The Hamiltonian of the Domain Wall Fermion}
The Hamiltonian of the free DWF can be expressed as
\begin{equation}
{\mathcal H}_1=\left(\begin{array}{cc}M^\dagger & -(\Slash{p}+\Slash{A})\\
                            (\Slash{p}+\Slash{A})& M\end{array}\right),
\end{equation}
and its square becomes
\begin{equation}
{{\mathcal H}_1}^\dagger{\mathcal H}_1=\left(\begin{array}{cc}M M^\dagger-(\Slash{p}+\Slash{A})^2& 0\\
0 & -(\Slash{p}+\Slash{A})^2+M^\dagger M\end{array}\right).
\end{equation}
Taking the eigenstates of hamiltonian including the gauge potential $\Slash{A}$ as the expansion bases and identifying
\begin{equation}
Q=\left(\begin{array}{cc} 0&-\Slash{p}\\
                          0&0\end{array}\right)
\qquad
Q^\dagger=\left(\begin{array}{cc} 0& 0\\
                          \Slash{p} &0\end{array}\right)
\end{equation}
as the supersymmetry operators that satisfy 
\[
Q^2={Q^\dagger}^2=0, \qquad \{Q,Q^\dagger\}=H
\]
and $[H,Q]=0$, we regard $M$ and $M^\dagger$ are 
 a pair of supersymmetric interactions \cite{NaNe93,CKS01}.

In the free fermionic theory, the number of massless right-handed particle is $dim(Ker(M))=n_R$ and the massless left-handed particle is $dim(Ker(M^\dagger))=n_L$.@It was shown that by choosing a proper operator $M$, one can define $U=M^\dagger(M M^\dagger)^{-1/2}$, such that $U^\dagger U=1$ and $UU^\dagger=M^\dagger(MM^\dagger)^{-1}M=1-Q$ where $Q$ is the projector on the zero eigenspace of $M$.

\end{document}